\documentclass[10 pt, conference]{ieeeconf}
\IEEEoverridecommandlockouts 
\makeatletter

\let\proof\@undefined
\let\endproof\@undefined
\let\labelindent\relax
\makeatother
\usepackage{eufrak}
\usepackage{cite}
\usepackage{hyperref}
\usepackage{dsfont}

\usepackage{xcolor}
\hypersetup{
	colorlinks,
	linkcolor={blue!70!black},
	citecolor={blue!70!black},
	urlcolor={blue!70!black}
}
\usepackage{amsmath}
\usepackage{hyphenat}
\usepackage{amsfonts}
\usepackage{amssymb,latexsym}
\usepackage[psamsfonts]{eucal}
\usepackage{amsthm}
\usepackage{graphicx}
\usepackage{epsfig}
\usepackage{epstopdf}
\usepackage{psfrag}
\usepackage{indentfirst}
\usepackage{tikz}
\usetikzlibrary{calc,patterns,decorations.pathmorphing,decorations.markings}
\usepackage{enumitem}
\usepackage{comment}
\usepackage{float}
\usepackage[font=small,skip=0pt]{caption}
\DeclareMathAlphabet{\bm}{U}{bbm}{m}{sl}

\allowdisplaybreaks

\renewcommand{\[}{\left[}

\usepackage{enumitem}

\usepackage[utf8]{inputenc}
\usepackage{graphicx}
\usepackage{BernsteinStyle}
\usepackage{latexsym,mathtools}
\usepackage[psamsfonts]{eucal}
\usepackage{fixltx2e} 
\usepackage{comment}
\usepackage{graphicx}
\usepackage{epsfig}
\usepackage{psfrag}
\usepackage{indentfirst}
\usepackage{epstopdf}
\usepackage{comment}
\usepackage{graphicx}
\usepackage[numbered]{bookmark}
\usepackage{float}
\usepackage[font=small,skip=0pt]{caption}

\newtheorem{theorem}{\indent Theorem}

\newtheorem{proposition}{\indent Proposition}

\allowdisplaybreaks

\newtheorem{lemma}{\indent Lemma}

\newtheorem{fact}{\indent Fact}

\newtheorem{conjecture}{\indent Conjecture}

\newtheorem{corollary}{\indent Corollary}

\newtheorem{definition}{\indent Definition}

\newtheorem{example}{\indent Example}

\newtheorem{experiment}{\indent Experiment}

\newcommand\xqed[1]{%
  \leavevmode\unskip\penalty9999 \hbox{}\nobreak\hfill
  \quad\hbox{#1}}
  
\newcommand\exampletriangle{\xqed{$\diamond$}}

\title{Output-Feedback Nonlinear Model Predictive Control \\ 
	with  Iterative State- and Control-Dependent Coefficients
}

\author{Mohammadreza Kamaldar \thanks{
		Department of Aerospace Engineering, University of Michigan, Ann Arbor,
		MI, USA.\ttfamily \{kamaldar,dsbaero\}@umich.edu}and Dennis S. Bernstein}

\begin{document}
	\maketitle

	\begin{abstract}
		By optimizing the predicted performance over a receding horizon, model predictive control (MPC) provides the ability to enforce state and control constraints.
		The present paper considers an extension of MPC for nonlinear systems that can be written in pseudo-linear form with state- and control-dependent coefficients.
		The main innovation is to apply quadratic programming iteratively over the horizon, where the predicted state trajectory is updated based on the updated control sequence.
		Output-feedback control is facilitated by using the block-observable canonical form for linear, time-varying dynamics.
		This control technique is illustrated on various numerical examples, including the Kapitza pendulum with slider-crank actuation, the nonholonomic integrator, the electromagnetically controlled oscillator, and the triple integrator with control-magnitude saturation.
	\end{abstract}
	

	
	\section{Introduction}
	
	By performing optimization over a future horizon, model predictive control (MPC) provides the means for controlling linear and nonlinear systems with state and control constraints
	\cite{gilbert1988,kwon2006receding,camacho2013model,eren2017model}.
	For systems with nonlinear dynamics, linearized plant models can be used iteratively to perform the receding-horizon optimization
	\cite{LiTodorov2004,TodorovLi2005}.
	Although convergence is not guaranteed, application of these techniques shows that they are effective in practice \cite{tomizukaiLQR}.

	The present paper develops a novel approach to MPC for nonlinear systems by performing the iterative receding-horizon optimization based on state- and control-dependent coefficients.
	By recasting the nonlinear dynamics as pseudo-linear dynamics, state- and control-dependent coefficients (SCDCs) provide a heuristic technique for nonlinear control.
	Although guarantees are limited to local stabilization \cite{alleyne}, SCDCs have seen widespread usage in diverse applications
	\cite{findeisen2003state,mracek1998control,erdem2004design,ccimen2010systematic,weiss2012forward,prach2018output,tsiotras1996counterexample}.

	In the prior literature, SCDCs have been used within the context of MPC for nonlinear systems.
	%
	In \cite{chang2013constrained}, a discrete-time SDRE technique is used with receding-horizon optimization based on the backward-propagating Riccati equation (BPRE) for enforcing state constraints.
	A related technique that fixes the current state over the future horizon is applied to quadcopter dynamics in \cite{ru2017nonlinear}.
	%

	%

	In the present paper, we describe iterative state- and control-dependent  MPC (ISCD-MPC)  for output-feedback control of nonlinear systems.
	In order to assess the performance of this technique, we consider  several well-studied benchmark problems, namely, the Kapitza pendulum \cite{bergAUTOMATICA2015,artstein2021,kapitzaACC23}, the nonholonomic integrator \cite{brockett1983asymptotic,bloch1992control,murray1993nonholonomic,kolmanovsky1995developments,bloch1996stabilization,hespanha1999stabilization,modin2020makes}, the electromagnetically controlled oscillator \cite{hong1998stabilization,cairano2007model,yan2012setpoint}, and the chain of integrators \cite{teel1992global,kamaldar2021dynamic}.
	The first three examples assume full-state feedback.
	For output-feedback control of the chain of integrators, we show how the state of the block-observable canonical form (BOCF) realization of the linear time-varying (LTV) pseudo-linear dynamics is an explicit function of the state and input matrices \cite{islamPCAC}.
	In effect, BOCF provides a deadbeat observer for the LTV pseudo-linear dynamics.

	As in all nonlinear control techniques, especially those based on heuristics, a fundamental challenge is to demonstrate acceptable performance over the largest possible domain of attraction.
	For the chain of integrators, we thus investigate the convergence of the state trajectory over a grid  of initial conditions for various values of the horizon.

	\section{Notation}
	%
	%
	%
	%
	%
	The symmetric matrix $P\in\BBR^{n\times n}$ is positive semidefinite (resp., positive definite) if all of its eigenvalues are nonnegative (resp., positive).
	Let $x_{(i)}$ denote the $i$th component of $x\in\BBR^n.$
	\section{Problem Formulation}
	Consider the discrete-time nonlinear  system
	\begin{equation} 
		x_{k+1}= f(x_k,u_k),
		\label{eq:xk1}
	\end{equation}
	where, for all $k\ge0$, $x_k\in\BBR^n$ is the state at step $k$,  $x_{0}\in\BBR^n$ is the fixed initial condition, $u_k\in\BBR^m$ is the control input applied at step $k$, $u_0\in\BBR^m$ is the given initial control input, and $f\colon \BBR^n\times \BBR^m\to\BBR^{n}.$ 
	Let $\ell\ge2$ be the horizon length, and, for all $j\in\{1,\ldots,\ell\}$, let $x_{k,j}$    be  the computed state for step $k+j$ obtained at step $k$ using
	\begin{align}
		x_{k,j+1} = f(  x_{k,j},  u_{k,j} ),
		\label{eq:xfka}
	\end{align}
	where $x_{k,0}\isdef x_k$, $u_{k,0}\isdef u_k$, and, for all $j\in\{1,\ldots,\ell-1\}$, $u_{k,j}$ is the computed control  for  step $k+j$ obtained at step $k.$
	Note that
	\begin{align}
		x_{k,1} = f( x_{k},u_{k} ).
		\label{eq:xk111x}
	\end{align}

	Next, for all $k\ge0,$ consider the performance measure
	\begin{align}
		J_k&(u_{k,1},\ldots,u_{k,\ell-1}) =  \tfrac{1}{2} x_{k,\ell}^\rmT Q_{k,\ell} x_{k,\ell}\nn\\
		&\quad +\tfrac{1}{2} \sum_{j=1}^{\ell-1} (x_{k,j}^\rmT Q_{k,j} x_{k,j} + u_{k,j}^\rmT R_{k,j} u_{k,j}) ,
		\label{Jdefn}
	\end{align} 
	where  
	$Q_{k,\ell}\in\BBR^{n\times n}$ is the positive-semidefinite terminal weighting, and, for all $j\in\{1,\ldots,\ell-1\},$  
	$Q_{k,j}\in\BBR^{n\times n}$ is the positive-semidefinite state weighting, and $R_{k,j}\in\BBR^{m\times m}$ is the positive-definite control weighting.

	At each time step $k\ge0$, the objective is to find a sequence of control inputs $u_{k,1},\ldots,u_{k,\ell-1}$ such that $J_k$ is minimized subject to \eqref{eq:xfka}, \eqref{eq:xk111x}, and the constraints
	\begin{gather}
		\SA v_k \le b,\\
		\SA_{\rm eq} v_k = b_{\rm eq},\\
		\underline v_s\le v_{k(s)}\le \overline v_s,~~~~~ s\in\{1,\ldots,\ell(n+m)-m\},
	\end{gather}
	where, $\SA,\SA_{\rm eq}\in\BBR^{n_\rmc\times (\ell(n+m)-m)}$, $b,b_{\rm eq} \in \BBR^{n_\rmc}$, $n_\rmc\ge0$ is the number of constraints, for all $k\ge0$,  $v_k \in\BBR^{\ell(n+m)-m}$ is defined by
	\begin{equation}
		v_k\isdef \matl x_{k,1}^\rmT & \cdots & x_{k,\ell}^\rmT&u_{k,1}^\rmT&\cdots&u_{k,\ell-1}^\rmT\matr^\rmT,
	\end{equation}
	and, for all $s\in\{1,\ldots,\ell(n+m)-m\},$ $\underline v_s,\overline v_s\in\BBR$ are such that $\underline v_s<\overline v_s.$  
	In accordance with  receding-horizon control, the first element $u_{k,1}$ of the sequence of computed controls is then applied  to the system at time step $k+1$, that is, for all $k\ge0,$
	\begin{equation}
		u_{k+1} =   u_{k,1},
	\end{equation}
	and $u_{k,2},\ldots,u_{k,\ell-1}$ are discarded.
	%
	The optimization is performed beginning at step $k$ and is assumed to be completed before step $k+1.$
	The optimization of \eqref{Jdefn} is performed by the iterative procedure detailed in the next section.

	\section{Iterative State- and Control-Dependent Model Predictive Control (ISCD-MPC)}

	We use a reformulation of the nonlinear system \eqref{eq:xk1}, namely, state- and control-dependent coefficients (SCDC) to present an iterative algorithm based on QP for computing $u_{k+1}$.

	Let $\rho\ge1$ denote the number of iterations, and let $i\in\{1,\ldots,\rho\}$ denote the index of the $i$th iteration at step $k.$ 
	For all $k\ge0$ and all $j\in\{1,\hdots,\ell\}$, let  $x_{k,j|i}\in\BBR^n$ denote  the  computed state for step $k+j$ obtained at time step $k$ and iteration $i$.
	Similarly, let
	$u_{k,j|i} \in\BBR^m$  denote  the computed control for step $k+j$ obtained at time step $k$  and iteration $i$.
	
	Let $A\colon \BBR^n\times \BBR^m\to \BBR^{n\times n}$ and $B\colon \BBR^n\times \BBR^m \to \BBR^{n\times m}$ be such that, for all $x\in\BBR^n$ and all $u\in\BBR^m,$
	\begin{equation}
		f(x,u) = A(x,u)x+B(x,u)u,
		\label{eq:fSCDC}
	\end{equation}
	which is an SCDC representation.
	For  all $j\in\{1,\ldots,\ell\}$, the computed state $  x_{k,j|i}$ is obtained using
	\begin{equation}
		x_{k,j+1|i}= A_{k,j|i} x_{k,j|i}+B_{k,j|i} u_{k,j|i},
		\label{eq:xkj+1i}
	\end{equation}
	where
	\begin{align}
		A_{k,j|i} \isdef A(x_{k,j|i},u_{k,j|i}),\quad B_{k,j|i} \isdef B(x_{k,j|i},u_{k,j|i}), 
		\label{eq:AB_SCDC}
	\end{align}
	and
	\begin{align}
		x_{k,0|i}&\isdef x_k,\\
		u_{k,0|i} &\isdef u_k.
		\label{eq:uk0i}
	\end{align}
	Note that \eqref{eq:fSCDC}--\eqref{eq:uk0i} implies that
	\begin{align}
		x_{k,1|i} &= f(x_k,u_k)\\
		&=A(x_k,u_k)x_k+B(x_k,u_k)u_k\\
		&=A_{k,0|i} x_k + B_{k,0|i} u_k.
	\end{align}

	Now, we use iterations of QP to compute $u_{k,j|i}$.
	For all $k\ge0$ and all $j \in\{ 1,\ldots,\ell-1\},$ initialize
	\begin{equation}
		u_{k,j|1} \isdef u_k. 
	\end{equation}
	%
	%
	%
	%
	For all $i\in\{2,\hdots,\rho\}$, let the computed control sequence $ \{u_{k,1|i},\ldots,u_{k,\ell-1|i}\}$  be the solution of the quadratic program
	\begin{align}
		\min_{\mu_{1},\ldots,\mu_{\ell-1}} \Bigg( 
		\tfrac{1}{2} \xi_{\ell}^\rmT Q_{k,\ell} \xi_{\ell}+\tfrac{1}{2} &\sum_{j=1}^{\ell-1} ( \xi_{j}^\rmT Q_{k,j} \xi_{j} + \mu_{j}^\rmT R_{k,j} \mu_{j}) \Bigg),
	\end{align}
	\mbox{subject to:~~~} 
	\begin{gather}
		\xi_{1}= x_{k,1|i},\\
		\xi_{j+1}= A_{k,j|i-1} \xi_{j}+B_{k,j|i-1} \mu_{j},\\
		\SA \nu \le b,\\
		\SA_{\rm eq}  \nu = b_{\rm eq},\\
		\underline v_s\le  \nu_{(s)}\le \overline v_s,~~~~~ s\in\{1,\ldots,\ell(n+m)-m\},
	\end{gather}
	where $\nu\in\BBR^{\ell(n+m)-m}$ is defined by
	\begin{equation}
		\nu \isdef  \matl \xi_{1}^\rmT& \cdots & \xi_{\ell}^\rmT&\mu_{1}^\rmT&\cdots&\mu_{\ell-1}^\rmT\matr^\rmT.  
	\end{equation}
	%
	Finally,  let 
	\begin{equation}
		u_{k+1}=u_{k,1|\rho}.
	\end{equation}

	\section{Stopping Criterion and Warm Starting}
	We present a modification of ISCD-MPC that can reduce the computational burden of the algorithm.
	In particular, at each step $k\ge0$, the modification uses a stopping criterion to potentially stop the iterations before reaching iteration $\rho$.
	Moreover, the modification uses the control sequence obtained at the last iteration of step $k$ to form the control sequence for the first iteration of step $k+1.$

	Let $\varepsilon>0$ be a tolerance used as the stopping criterion.
	For all $k\ge0$ and all $i\in\{1,\ldots,\rho\},$ define
	\begin{equation}
		U_{k|i} \isdef \matl u_{k,1|i}^\rmT &\cdots &u_{k,\ell-1|i}^\rmT \matr^\rmT \in\BBR^{m(\ell-1)}.
		\label{eq:UUUk}
	\end{equation}
	For each $k\ge0,$  define $\rho_k\in\{2,\ldots,\rho\}$  by
	\begin{equation}
		\rho_k\isdef \min\{\rho,\min\{i\in\{2,\ldots,\rho\}\colon \|U_{k|i}-U_{k|i-1}\|<\varepsilon\}\},
		\label{eq:rhokk}
	\end{equation}
	and let $\rho_k$ be the index of the last iteration.
	For  all $j \in\{ 1,\ldots,\ell-1\},$ initialize
	\begin{equation}
		u_{0,j|1} \isdef u_0, 
		\label{eq_u0j1}
	\end{equation}
	and, for all $k\ge1$ and all $j \in\{ 1,\ldots,\ell-1\},$   initialize
	\begin{align}
		u_{k,j|1} &\isdef \begin{cases}
			u_{k-1,j+1|\rho_{k-1}}, & j\in\{1,\ldots,\ell-2\},\\
			u_{k-1,\ell-1|\rho_{k-1}}, & j=\ell-1.
		\end{cases} 
		\label{eq_ukj33}
	\end{align}
	
	For all examples in this paper, we use ISCD-MPC with the stopping criteria defined by \eqref{eq:UUUk} and \eqref{eq:rhokk}, and the warm starting defined by \eqref{eq_u0j1} and \eqref{eq_ukj33}.

	Each example in this paper includes magnitude saturation of the control input.
	This constraint can be enforced directly using QP.
	In the present paper, however, we represent the magnitude saturation $\sigma(u)$ as $[\sigma(u)/u]u,$ so that $B\sigma(u)/u$ is treated as a control-dependent coefficient.
	Numerical experience shows that this technique is more reliable than direct enforcement of control constraints using QP.


	The input nonlinearity $\sigma$ has the property that $\sigma(u)/u$ has a removable singularity at $u=0$ with the value $\sigma(0)/0\isdef \lim_{u\to0}\sigma(u)/u.$
	Similarly, $\sin(0)/0\isdef1.$
	
	All examples in this paper are performed using a sampled-data control setting.
	In particular, MATLAB `ode45' command is used to simulate the continuous-time, nonlinear dynamics, where the `ode45' relative and absolute tolerances are set to $10^{-5}.$
	In addition, for all examples, we use MATLAB `quadprog' command to perform QP, where we choose $Q_{k,j}$ and $R_{k,j}$ to be independent of $k$ and $j$, and we thus write $Q$ and $R$.

\begin{figure}[t!]
		\centering
		\vspace{5pt}\includegraphics{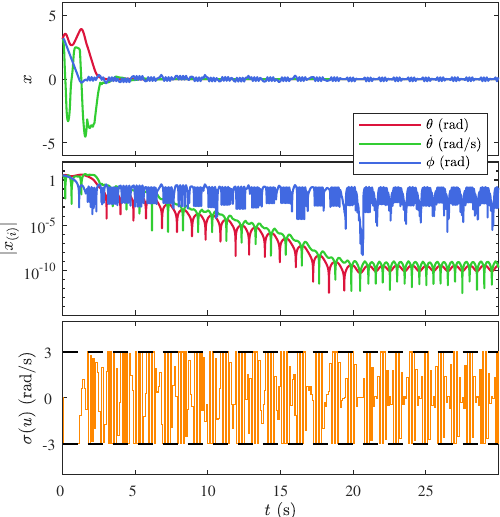}
		\caption{Example \ref{ex:K_pendulum}. Stabilization of Kapitza pendulum with slider-crank actuation subject to control-magnitude saturation.}
		\label{fig:K_pend}
	\end{figure}

	\begin{example}
		\label{ex:K_pendulum}
		Kapitza pendulum with slider-crank actuation subject to control-magnitude saturation.
		\rm
		Consider the Kapitza pendulum shown in Figure \ref{fig:Kapitza_pend_slider}, where the control input $u$ is the angular wheel speed.
		Assuming for simplicity that $(\sin\phi)^2 \ll (a/r)^2,$ the equations of motion are given by
		\begin{gather}
			\ddot\theta - 
			\frac{g}{l}\sin\theta +\frac{r}{l}\left(\cos \phi + \frac{r\cos 2\phi}{a}\right)(\sin\theta) \left(\sigma(u)\right)^2=0, \label{eq:Jth24e}\\
			\dot\phi = \sigma(u),\label{eq:htheta}
		\end{gather}
		%
		%
		where $ r$ is the radius of the wheel, $a$ is the length of the arm, $l$ is the length of the pendulum, $g$ is the acceleration of gravity, and $\sigma\colon \BBR\to\BBR$ is the control-magnitude saturation function defined by
		\begin{equation}
			\sigma(u)\isdef \begin{cases} u_{\max},&u> u_{\max},\\
				u,& u_{\min}\le u \le u_{\max},\\
				u_{\min}, & u<u_{\min},\end{cases}
			\label{eq:sat1}
		\end{equation}
		and $u_{\min},u_{\max}\in\BBR$ are the lower and upper magnitude saturation levels. 
		
		\begin{figure}[t!]
			\centering
			\vspace{2pt}
			\scalebox{1}{\includegraphics{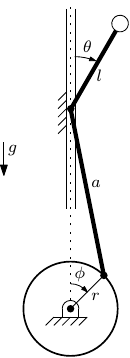}}
			\caption{Example \ref{ex:K_pendulum}. Kapitza pendulum with slider-crank actuation. The base of the pendulum, which is confined to a vertical slot, is actuated by an arm of length $a$ attached to the circumference of the wheel of radius $r.$}
			\label{fig:Kapitza_pend_slider}
		\end{figure}
		
		Let $x\isdef \matl \theta & \dot\theta&\phi \matr^\rmT$ be the state, and it follows from \eqref{eq:Jth24e} and \eqref{eq:htheta} that
		\begin{align}
			\dot x & =  \matl x_{(2)} \\ \frac{g}{l}\sin x_{(1)}\\0\matr \nn\\
			&\quad + \matl 0 \\ -\frac{r}{l}\left(\cos x_{(3)} + \frac{r\cos 2x_{(3)}}{a}\right)(\sin x_{(1)})(\sigma(u))^2\\\sigma(u)\matr.
			\label{eq:kap_pend}
		\end{align}
		For all $k\ge0$ and all $t\in[kT_\rms,(k+1)T_\rms)$, we consider $u(t) =u_k,$ 
		%
		where $u_k\in\BBR$ is the discrete-time control determined by the algorithm.  
		%
		%
		Using Euler integration, the discrete-time approximation of \eqref{eq:kap_pend} is given by
		\begin{align}
			&x_{k+1} = x_k +T_\rms  \matl x_{k{(2)}} \\ \frac{g}{l}\sin x_{k{(1)}}\\0\matr \nn\\
			&+T_\rms  \matl 0 \\ -\frac{r}{l}\left(\cos x_{k{(3)}} + \frac{r\cos 2x_{k{(3)}}}{a}\right)(\sin x_{k{(1)}})(\sigma(u_k))^2\\\sigma(u_k)\matr,
			\label{eq:xk1bbb44}
		\end{align}
		where, for all $k\ge0,$ $x_k\isdef x(kT_\rms)$.
		To implement ISCD-MPC, we consider the SCDC representation of \eqref{eq:xk1bbb44} given by
		\begin{align}
			x_{k+1} &= x_k + T_\rms \matl 0 &1 &0 \\ \frac{g\sin x_{k{(1)}}}{lx_{k{(1)}}}&0&0\\0&0&0\matr x_k\nn\\
			&\quad + T_\rms \matl 0 \\  \frac{-\frac{r}{l}\left(\cos x_{k(3)} + \frac{r\cos 2x_{k(3)}}{a}\right)(\sin x_{k(1)})(\sigma(u_k))^2}{u_k} \\[1ex] \frac{\sigma(u_k)}{u_k}
			\matr u_k.
		\end{align}
		Figure \ref{fig:K_pend} shows $x$ and $\sigma(u)$ using ISCD-MPC, where
		\begin{gather}
			T_\rms=0.1\,\rms,\quad   l=0.25\,\rmm,\quad r=1\,\rmm,\\
			a=2\,\rmm,\quad g=9.81\,\tfrac{\rmm}{\rms^2},\\ x_0=\matl\pi\,{\rm rad}& \pi\,\tfrac{\rm rad}{\rms}&\pi\,{\rm rad}\matr^\rmT,\quad u_0=0,\\  u_{\max} =-u_{\min}=3\,\tfrac{\rm rad}{\rms},\quad
			\ell=50,\quad  \rho=30,\\ \varepsilon=10^{-3},\quad  Q={\rm diag}(10^4,10^3,10^6),\quad  R=1.  \exampletriangle
		\end{gather}
	\end{example}
	
\begin{figure}
		\centering
		\vspace{5pt}    \includegraphics{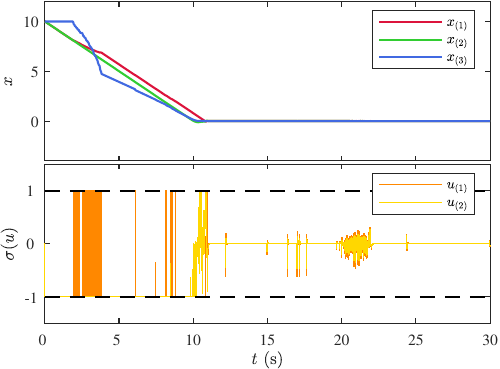}
		\caption{Example \ref{ex_NH_integrator}. Stabilization of the nonholonomic integrator subject to control-magnitude saturation.}
		\label{fig:NH_integrator}
	\end{figure}

	\begin{example}
		\label{ex_NH_integrator}
		Nonholonomic integrator with control-magnitude saturation. \rm
		Consider the continuous-time nonholonomic integrator with control-magnitude saturation given by
		\begin{equation}
			\dot x = \matl 1 &0 \\ 0 &1\\-x_{(2)} & x_{(1)}\matr \sigma(u),
		\end{equation}
		where $\sigma\colon\BBR^2\to\BBR^2$ is the control-magnitude saturation function  
		\begin{equation}
			\sigma(u) \isdef  \matl \bar \sigma(u_{(1)})\\\bar\sigma(u_{(2)})\matr,
			\label{eq:Hsat1}
		\end{equation}
		where $\bar\sigma\colon \BBR\to\BBR$ is defined by 
		\begin{equation}
			\bar\sigma(u)\isdef \begin{cases} u_{\max},&u> u_{\max},\\
				u,& u_{\min}\le u \le u_{\max},\\
				u_{\min}, & u<u_{\min},\end{cases}
		\end{equation}
		and $u_{\min},u_{\max}\in\BBR$ are the lower and upper magnitude saturation levels, respectively.
		Let $T_\rms>0$, and, for all $k\ge0$ and all $t\in[kT_\rms,(k+1)T_\rms)$,  consider $u(t) =u_k,$  
			%
		%
		where $u_k\in\BBR$ is the discrete-time control determined by the algorithm.  
		Using Euler integration, the discrete-time approximation of \eqref{eq:kap_pend} is given by
		\begin{equation}
			x_{k+1} = x_k +T_\rms \matl1&0\\0&1\\-x_{k{(2)}} & x_{k{(1)}}\matr \sigma(u_k),
			\label{eq:nhdt}
		\end{equation}
		where, for all $k\ge0,$ $x_k\isdef x(kT_\rms)$.
		To implement ISCD-MPC with control magnitude saturation, we consider the SCDC representation of \eqref{eq:nhdt} given by
		\begin{equation}
			x_{k+1} = x_k + B(x_k,u_k) u_k,
		\end{equation}
		where
		\begin{equation}
			B(x_k,u_k) = T_\rms \matl1&0\\0&1\\-x_{k{(2)}} & x_{k{(1)}}\matr\cdot \begin{cases} 
				I_2,&  {\sigma(u_k)=u_k},\\
				\frac{\sigma(u_k)u_k^\rmT}{\|u_k\|^2}, & {\sigma(u_k)\ne u_k}.
			\end{cases}
		\end{equation}
		Figure \ref{fig:NH_integrator}  shows $x$ and $\sigma(u)$ using ISCD-MPC, where
		\begin{gather}
			T_\rms=0.01\,\rms,\quad x_0=\matl10& 10&10\matr^\rmT,\\ u_0=0,\quad  u_{\max} =-u_{\min}=1,\quad
			\ell=500,\quad  \rho=50,\\ \varepsilon=10^{-3},\quad  Q={\rm diag}(10^3,10^3,10^4),\quad  R=I_2.\exampletriangle  %
		\end{gather}
		
	\end{example}
	
\begin{figure}[t!]
		\centering
		\vspace{5pt}\includegraphics{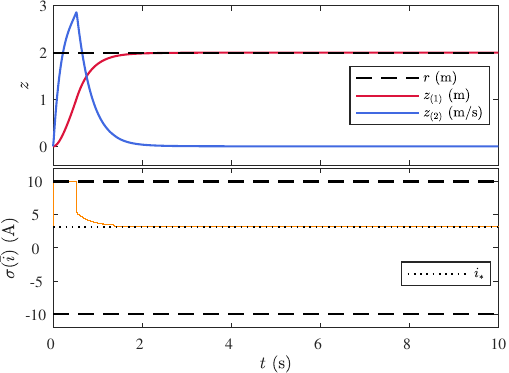}
		\caption{Example \ref{ex_electro_mag}. Command following for the electromagnetically controlled oscillator subject to control-magnitude saturation.}
		\label{fig:electro_mag}
	\end{figure}

	\begin{example}
		\label{ex_electro_mag}
		Electromagnetically controlled oscillator.
		\rm
		Consider the electromagnetically controlled oscillator shown in Figure \ref{fig:elec_mag}, where $q$ is the position of the mass $\bar m$, $\bar k$ is the stiffness of the spring, $\bar b$ is damping, $\bar q$ is the distance from mass to the magnet when the spring is relaxed, and $\bar i$ is the current.
		Define $z\isdef \matl q&\dot q\matr^\rmT$.
		Then, the governing dynamics of the electromagnetically controlled oscillator are  given by
		\begin{equation}
			\dot z = \matl z_{(2)} \\ -\frac{\bar b}{\bar m} z_{(2)} - \frac{\bar k}{\bar m} z_{(1)} + \frac{\bar\varepsilon (\sigma(\bar i))^2}{m(\bar q-z_{(1)})^2} \matr,
			\label{eq:dzz2}
		\end{equation}
		where $\bar\varepsilon$ is a force constant representing the strength of the electromagnet, and $\sigma\colon \BBR\to\BBR$ is the control-magnitude saturation function defined by \eqref{eq:sat1}.
		\begin{figure}[b!]
			\centering
			\scalebox{1}{  \includegraphics{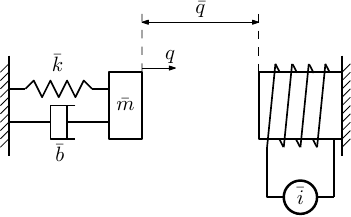}}
			\caption{Electromagnetically controlled oscillator}
			\label{fig:elec_mag}
		\end{figure}
		Let $r\in(0,\bar q)$ be the desired position of the mass, and define $x\isdef z-\matl r& 0\matr^\rmT,$ which implies $\dot x=\dot z.$
		It thus follows from \eqref{eq:dzz2} that 
		\begin{equation}
			\dot x = \matl x_{(2)} \\ -\frac{\bar b}{\bar m} x_{(2)} - \frac{\bar k}{\bar m} (x_{(1)}+r) + \frac{\bar\varepsilon (\sigma(\bar i))^2}{\bar m(\bar q-x_{(1)}-r)^2} \matr.
			\label{eq:dzz24}
		\end{equation}
		Let $T_\rms>0$, and, for all $k\ge0$ and all $t\in[kT_\rms,(k+1)T_\rms)$, consider $\bar i(t) =\bar i_k,$  
		%
		%
		where $\bar i_k\in\BBR$ is the discrete-time current determined by the algorithm.  
		Using Euler integration, the discrete-time approximation of \eqref{eq:dzz24} is given by
		\begin{align}
			x_{k+1} &= x_k +T_\rms  \matl x_{k{(2)}} \\ -\frac{\bar b  x_{k{(2)}}}{\bar m} - \frac{\bar k x_{k{(1)}}}{\bar m}  + \frac{\bar\varepsilon (\sigma(\bar i_k))^2}{\bar m(\bar q-x_{k{(1)}}-r)^2}-\frac{kr}{m} \matr\nn\\
			& = x_k +T_\rms  \matl x_{k{(2)}} \\ -\frac{\bar b  x_{k{(2)}}}{\bar m} - \frac{\bar k x_{k{(1)}}}{\bar m}  + \frac{\bar\varepsilon (\sigma_{\rma}(u_k))^2}{\bar m(\bar q-x_{k{(1)}}-r)^2}  \matr, 
			\label{eq:nhdt31}
		\end{align}
		where, for all $k\ge0,$ $x_k\isdef x(kT_\rms)$, 
		\begin{equation}
			u_k\isdef \bar i_k -\bar i_*,
		\end{equation}
		where $\bar i_*\isdef \sqrt{\frac{(\bar q-r)^2\bar k r}{\bar \varepsilon}}$ is the equilibrium current,  and $\sigma_\rma\colon\BBR\to\BBR$ is the asymmetric control-magnitude saturation function defined by
		\begin{equation}
			\sigma_\rma(u)\isdef \begin{cases} u_{\max}-\bar i_*,&u> u_{\max}-\bar i_*,\\
				u,& u_{\min}-\bar i_*\le u \le u_{\max}-\bar i_*,\\
				u_{\min}-\bar i_*, & u<u_{\min}-\bar i_*.\end{cases}
		\end{equation}
		To implement ISCD-MPC with control magnitude saturation, we consider the SCDC representation of \eqref{eq:nhdt31} given by
		\begin{align}
			x_{k+1} &= x_k +T_\rms  \matl 0 & 1\\ - \frac{\bar k}{\bar m}  & -\frac{\bar b}{\bar m} \matr x_k\nn\\
			&\quad + T_\rms\matl 0 \\ \frac{\bar\varepsilon (\sigma_\rma(u_k))^2}{\bar m u_k(\bar q-x_{k{(1)}}-r)^2}\matr u_k.
		\end{align}
		Figure \ref{fig:electro_mag}  shows $z$ and $\sigma(\bar i)$ using ISCD-MPC, where
		\begin{gather}
			T_\rms=0.01\,\rms,\quad z(0)=\matl0{\,\rm m}& 0{\,\rm m/s}\matr^\rmT,\\
			r=2{\,\rm m},\quad \bar q=3{\,\rm m}, \quad \bar\varepsilon=1{\,\rm N.m^2.A^{-2}} \\
			\bar k=5{\,\rm N/ m},\quad \bar m=1{\,\rm kg},\quad \bar c=5{\,\rm N.s/m},\\ 
			u_0=10^{-2}{\,\rm A},\quad  u_{\max} =-u_{\min}=10{\,\rm A},\\
			\ell=300,\quad  \rho=50,\quad \varepsilon=10^{-3},\\  Q={\rm diag}(10^3,10^{2}),\quad  R=1.  %
		\end{gather}
		Note that the current $\bar i$  converges to $\bar i_*=\sqrt{10}$~A.
		\exampletriangle
	\end{example}
	
	\section{Output-Feedback Control of  Input-Output Systems}
	
	Consider the discrete-time input-output system 
	\begin{align}
		y_{k} &=-\sum_{\tau=1}^n f_\tau(y_{k-1},\ldots,y_{k-n},u_{k-1},\ldots, u_{k-n})y_{k-\tau} \nn\\
		&\quad + \sum_{\tau=1}^n g_\tau(y_{k-1},\ldots,y_{k-n},u_{k-1},\ldots, u_{k-n})u_{k-\tau}, \label{eq:yyy}
	\end{align}
	where, for all $\tau\in\{1,\ldots,n\}$,  $f_\tau\colon \BBR^p \times \cdots \times \BBR^p\times \BBR^m \times \cdots \times \BBR^m\to\BBR^{p\times p}$ and  $g_\tau\colon \BBR^p \times \cdots \times \BBR^p\times \BBR^m \times \cdots \times \BBR^m \to\BBR^{p\times m},$
	and, for all  $k\ge0,$ $y_k\in\BBR^p$ is the measured output, $u_k\in\BBR^m$ is the control input, and 
	\begin{gather}
		y_{-1}=\cdots=y_{-n}=0,\\ u_{-1}=\cdots=u_{-n}=0.
	\end{gather}
	For all $\tau\in\{1,\ldots,n\}$ and all $k\ge0$, define
	\begin{align}
		F_{\tau,k} &\isdef f_\tau(y_{k-1},\ldots,y_{k-n},u_{k-1},\ldots, u_{k-n})\in\BBR^{p\times p}, \\
		G_{\tau,k} &\isdef g_\tau(y_{k-1},\ldots,y_{k-n},u_{k-1},\ldots, u_{k-n})\in\BBR^{p\times m}.
	\end{align}
	%
	%
	%
	Then, for all $k\ge0$, the block observable canonical form (BOCF) of \eqref{eq:yyy} is given by \cite{polderman1989state}
	\begin{align}
		x_{k+1} &= A(x_k,u_k)x_{k} + B(x_k,u_k) u_{k},\label{eq:xk1_OF}\\
		y_{k} & = C x_{k},
	\end{align}
	where
	\begin{gather}
		A(x_k,u_k) \isdef \matl -F_{1,k+1} & 1 & 0 & \cdots & 0  \\
		-F_{2,k+1} & 0 & 1 &\cdots  & 0\\
		\vdots & \vdots & \vdots & \ddots & \vdots \\
		-F_{3,k+1} & 0 & 0 & \cdots & 1\\
		-F_{n,k+1} & 0 & 0 & \cdots & 0
		\matr \in \BBR^{np\times np},       \\ 
		B(x_k,u_k) \isdef \matl  G_{1,k+1}  \\
		G_{2,k+1}  \\
		\vdots   \\
		G_{n,k+1} 
		\matr  \in\BBR^{np\times m},\label{eq:AoBo}\\
		C \isdef \matl I_p & 0 & \cdots & 0 & 0\matr \in \BBR^{p\times np},
		\label{eq:CD}
	\end{gather}
	and 
	\begin{equation}
		x_k\isdef \matl x_{k(1)} \\  \vdots\\ x_{k(n)}\matr \in\BBR^{np},
	\end{equation}
	where 
	\begin{gather}
		x_{k(1)} \isdef y_k,
		\label{eq:eta1}
	\end{gather}
	and, for all $\tau\in\{ 2,\hdots,n\},$ $x_{k(\tau)}\in\BBR^p$ is defined by 
	\begin{equation}
		x_{k(\tau)} \isdef \sum_{s=1}^{n-\tau+1}  -F_{\tau+s-1,k} y_{k-s}  +  \sum_{s=1}^{n-\tau+1}  G_{\tau+s-1,k} u_{k-s}.
		\label{eq:eta2}
	\end{equation}
	
	Since \eqref{eq:xk1_OF} represents pseudo-linear dynamics of the nonlinear system \eqref{eq:yyy}  using SCDC, we can apply ISCD-MPC, as demonstrated in the following example. 
	
\begin{figure}[t!]
		\centering
		\vspace{5pt}\includegraphics{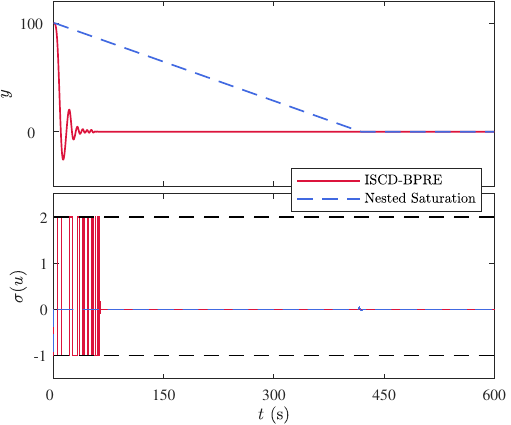}
		\caption{Example \ref{ex:triple_int}. Comparison of output-feedback stabilization of a triple integrator subject to asymmetric control-magnitude saturation using ISCD-MPC with the nested-saturation controller of \cite{kamaldar2021dynamic}.}
		\label{fig:triple_int}
	\end{figure}

	\begin{example}
		\label{ex:triple_int}
		Triple-integrator with asymmetric control-magnitude saturation.
		\rm
		Consider the continuous-time system
		\begin{align}
			\dot x(t) &= \matl0& 1&0\\0 &0 &1\\0&0&0\matr x(t) + \matl 0\\0\\ 1\matr \sigma (u(t)),\label{eq:xdot1}\\
			y(t) & =\matl 1 & 0 & 0\matr x,\label{eq:yt2}
		\end{align}
		where, for all $t\ge0$, $x(t)\in\BBR^3$ is the state, $u(t)\in\BBR$ is the control input, $y(t)\in\BBR$ is the measured output, and $\sigma\colon\BBR\to\BBR$ is the control-magnitude saturation function defined by \eqref{eq:sat1}.
		If, for all $t\ge0$, $u(t)\in[u_{\min},u_{\max}]$, then \eqref{eq:xdot1} and \eqref{eq:yt2} yield the transfer function
		\begin{equation}
			G_\rmc(s) = \frac{1}{s^3},\label{eq:Gc}
		\end{equation}
		which is a triple integrator.
		Let $T_\rms=0.1$~s be the sample time, and, for all $k\ge0$,
		define $y_k \isdef y(kT_\rms)$. 
		Moreover, for all $k\ge0$ and all $t\in [kT_\rms, (k+1)T_\rms)$,
		let $u(t) = u_k$.
		The discrete-time counterpart of \eqref{eq:Gc}, which is the
		transfer function from $u_k$ to $y_k$, is given by
		\begin{equation}
			G(\bfq) = \frac{5(\bfq^2+ 4\bfq+1)}{3\times 10^4(\bfq-1)^3},\label{eq:Gqq}
		\end{equation}
		where $\bfq$ is the forward-shift operator.
		Note that the zeros of \eqref{eq:Gqq} are sampling zeros, and that $G$ is nonminimum phase \cite{aastrom2013computer}.
		Using \eqref{eq:Gqq}, the discrete-time input-output counterpart of \eqref{eq:xdot1} and \eqref{eq:yt2} is given by
		\begin{align}
			y_k &= 3 y_{k-1} -3y_{k-2}+y_{k-3} \nn\\
			&\quad + \tfrac{5}{3\times 10^{4}} \big(\sigma(u_{k-1}) +  4\sigma(u_{k-2}) + \sigma(u_{k-3})\big),
		\end{align}
		which has the form of \eqref{eq:yyy}, where
		\begin{gather}
			F_{1,k}= 3,\quad F_{2,k} = -3,\quad
			F_{3,k} = 1,\\
			G_{1,k}  = \tfrac{5\sigma(u_{k-1})}{3\times 10^{4}u_{k-1}},
			\quad
			G_{2,k} =  \tfrac{20\sigma(u_{k-2})}{3\times 10^{4}u_{k-2}},\\
			G_{3,k} = \tfrac{5\sigma(u_{k-3})}{3\times 10^{4}u_{k-3}}.
			%
		\end{gather}
		The red curves  in Figure \ref{fig:triple_int} shows $x$ and $\sigma(u)$ using ISCD-MPC, where 
		\begin{gather}
			x(0)=\matl300&0&0\matr^\rmT,\quad \ell=200,\quad \rho=30,\\ \varepsilon=10^{-3},\quad Q=10^{10}I_3,\quad R=1,\\ u_{\max} =2,\quad u_{\min} = -1.
		\end{gather}

		Next, we compare the performance of ISCD-MPC and the nested-saturation controller of \cite{teel1992global}.
		We consider the output-feedback version of \cite{teel1992global} presented in \cite{kamaldar2021dynamic}.
		Note that  \cite{teel1992global,kamaldar2021dynamic} are not predictive, whereas ISCD-MPC is a predictive algorithm.
		The blue curves  in Figure \ref{fig:triple_int} shows $x$ and $\sigma(u)$ with the nested-saturation controller of \cite{kamaldar2021dynamic}, where
		\begin{gather}
			u_0=0,\quad \lambda_1=\lambda_2=\lambda_3=-0.1,\\
			\overline\varepsilon_1=-\underline\varepsilon_1= 0.24 ,\quad \overline\varepsilon_2=-\underline\varepsilon_2= 0.25 ,\\\underline\varepsilon_2= -0.51,\quad \overline\varepsilon_3=1.51.\exampletriangle
		\end{gather}
	\end{example}
	
	\begin{figure}[b!]
		\centering
		\includegraphics{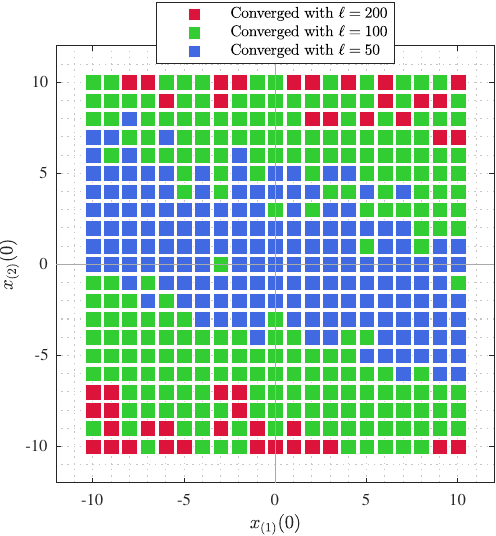}
		\caption{Example \ref{ex_triple_integrator_DOA}. Domain of attraction for the triple integrator \eqref{eq:Gc} subject to asymmetric control-magnitude saturation, where the initial conditions are $x_{(1)}(0),x_{(2)}(0)\in[-10,10]$ and $x_{(3)}(0)=0.$ 
			Note that the domain of attraction becomes larger as the horizon $\ell$ of ISCD-MPC increases.}
		\label{fig_triple_integrator_DOA}
	\end{figure}

	\begin{example}
		\label{ex_triple_integrator_DOA}
		Domain of attraction for a triple integrator with asymmetric control-magnitude saturation.  \rm
		For the chain of integrators in Example \ref{ex:triple_int}, we investigate the domain of attraction using ISCD-MPC.
		In particular, we consider the grid of initial conditions where $x_{(1)}(0),x_{(2)}(0)\in\{-10,9,\ldots,9,10\}$ and  $x_{(3)}(0) = 0.$ 
		Each simulation is run for $60$~s, which, since $T_\rms=0.1$~s, yields 600 step. 
		The convergence criterion is $\sum_{k=580}^{600}\|x_k\|<0.01.$
		Figure \ref{fig_triple_integrator_DOA} shows the domain of attraction of ISCD-MPC for  $\ell\in\{50,100,200\}$.
		%
		%
		Numerical simulations with larger values of $\ell$ (not shown) suggest that ISCD-MPC provides semiglobal stabilization. 
		\exampletriangle
	\end{example}

	\section{Conclusions}
	This paper presented ISCD-MPC, which is a model predictive control technique for full-state- or output-feedback control of nonlinear systems.
	In order to address the nonlinear dynamics, a state- and control-dependent parameterization of the nonlinearities was used.
	The nonlinearities were handled by means of an iterative technique involving QP.
	This technique was illustrated by means of a collection of nonlinear benchmark problems.

	A fundamental requirement of this technique is the need for the state- and control-dependent iteration of QP to converge at each step.
	Numerically, this iteration was found to be extremely reliable, a property shared with the iLQR technique \cite{TodorovLi2005,tomizukaiLQR}.
	A deeper understanding of the mechanisms responsible for convergence warrants further investigation.

	Several extensions of ISCD-MPC can be considered.
	%
	%
	Further numerical studies are required to understand the relationship between the achievable domain of attraction and the required horizon length \cite{worthmann2012estimates}, as well as the choice of the control- and state-dependent coefficient.
	In particular, based on Example \ref{ex_triple_integrator_DOA}, we conjecture that, with sufficiently large horizon, ISCD-MPC semiglobally stabilizes chains of integrators of arbitrary length with arbitrary zeros.

	\bibliographystyle{IEEEtran}
	\bibliography{Ref,MPCrefs,Kapitzarefs,SDCpapers}

\end{document}